\documentclass[letters,a4paper,fleqn,usenatbib]{mnras}

\usepackage{amsmath}	
\usepackage{amssymb}	
\usepackage{graphicx}	

\usepackage[T1]{fontenc}
\usepackage{ae,aecompl}

\usepackage{graphicx,times}
\usepackage{colortbl}
\usepackage{soul,color}
\usepackage[normalem]{ulem}
\usepackage{caption}
\usepackage{upgreek}
\usepackage{times,txfonts}

\def\aap{{A\&A}}		
 
\def\apj{{ApJ}}
\def\apjl{{ApJL}}		
\def\apjs{{ApJS}}		
\def\aj{{AJ}}
\def\apss{{Ap\&SS}} 
\def\mnras{{MNRAS}}

\def\ltorder{\mathrel{\raise.3ex\hbox{$<$}\mkern-14mu
             \lower0.6ex\hbox{$\sim$}}}

\def\r1{$r_1$}

\title[Coalescing Binary Black Holes Originating From Globular Clusters]{MOCCA-SURVEY Database I: Coalescing Binary Black Holes Originating From Globular Clusters}
\author[A. Askar,  M. Szkudlarek, D. Gondek-Rosi{\'n}ska, M. Giersz \& T. Bulik]{Abbas Askar$^{1}$\thanks{E-mail: askar@camk.edu.pl (AA)}, Magdalena Szkudlarek$^{2}$, Dorota Gondek-Rosi{\'n}ska$^{2}$,\newauthor Mirek Giersz$^{1}$ and Tomasz Bulik$^{3}$\\
$^{1}$Nicolaus Copernicus Astronomical Centre, Polish Academy of Sciences, 
ul. Bartycka 18, 00-716 Warsaw, Poland\\
$^{2}$Janusz Gil Institute of Astronomy, University of Zielona G{\'o}ra, Licealna 9, 65-407 Zielona G{\'o}ra, Poland\\
$^{3}$Astronomical Observatory Warsaw University, 00-478 Warsaw, Poland\\
\vspace{-0.6cm}}

\date{Accepted XXX. Received YYY; in original form ZZZ}

\pubyear{2016}

\begin{document}
\label{firstpage}
\pagerange{\pageref{firstpage}--\pageref{lastpage}}
\maketitle

\begin{abstract}
In this first of a series of papers, we utilize results for around two thousand star cluster models simulated using the \textsc{mocca} code for star cluster evolution (Survey I Database) to determine the astrophysical properties and local merger rate densities for coalescing binary black holes (BBHs) originating from globular clusters (GCs). We extracted information for all coalescing BBHs that escape the cluster models and subsequently merge within a Hubble time along with BBHs that are retained in our GC models and merge inside the cluster via gravitational wave (GW) emission. By obtaining results from a substantial number of realistic star cluster models that cover different initial parameters, we have an extremely large statistical sample of BBHs with stellar mass and massive stellar BH ($\lesssim 100M_{\odot}$) components that merge within a Hubble time. Using this data, we estimate local merger rate densities for these BBHs originating from GCs to be at least 5.4 ${\rm Gpc}^{-3}\,{\rm yr}^{-1}$.    
\end{abstract}



\section{Introduction}\label{sec:int}

The direct detection of the first GWs from a BBH merger \citep{GW1-2016} by the two detectors of the advanced Laser Interferometer Gravitational-Wave Observatory (LIGO) has not only confirmed the presence of GWs but has also ushered astrophysics into a new era of observing cosmic events that were previously invisible. Following the detection of GW150914 \citep{GW1-Astro}, subsequently one more confirmed BBH merger, GW151226 was detected by LIGO along with the detection of a BBH merger candidate LVT151012 \citep{GW2-2016}. There are significant astrophysical implications for the detection of these events. Firstly, these detections confirm the existence of coalescing BBH systems. Secondly, the masses of $\sim 30M_{\odot}$ for the BHs inferred from GW150914 \citep{GW1-Astro} show that massive stellar BHs ($\lesssim 100M_{\odot}$) do exist in the Universe. The formation scenario for massive BBHs and the origin of the detected coalescing binaries remains debatable. Such systems may form in the field via binary stellar evolution or they could have a dynamical origin in dense stellar environments like GCs or galactic nuclei \citep{ant2,ant1}. It is also possible that the detected events maybe coalescing primordial BHs \citep{sasaki}.

Dynamical processes in dense environments like GCs are conducive to forming BBHs \citep{SH1993}. Furthermore, most GCs have very low metallicities which can result in the formation of massive stellar BHs \citep{KB2010} that may form BBHs due to mass segregation during cluster evolution. Significant work has been done in using numerical simulations of GCs to predict the detection rates for GW events. Monte-Carlo codes for star cluster evolution have been used by \citet[]{Downing2011} to estimate yearly detection rates for GW radiation. More recently, \citet{Rodriguez1,Rodriguez2} have utilized Monte-Carlo simulations of about fifty GC models with updated treatment of stellar evolution for massive stars to estimate merger rates and properties of BBHs originating from GCs. Results from limited N-body simulations have also been used to estimate GW merger rates in recent studies by \citet{BSK2010,Tanikawa2013,BKL2014,Mapelli2016}.

\textbf{\begin{table*}
\begin{minipage}{183mm}
  \begin{center}
  \caption{Initial conditions for the Survey I models simulated using the \textsc{mocca} code.}
  \label{tab:models}
\begin{tabular}{|c|c|c|c|c|c|c|c|}
\hline
\textbf{N}      & \textbf{\begin{tabular}[c]{@{}c@{}}$R_{t}$\\ (pc)\end{tabular}} & \textbf{$R_t/R_h$} & \textbf{$W_0$} & \textbf{Z}                                                               & \textbf{$f_b$}       & \textbf{\begin{tabular}[c]{@{}c@{}}Natal Kicks\\  (km/s)\end{tabular}} & \textbf{\begin{tabular}[c]{@{}c@{}}Number of\\  Models\end{tabular}} \\ \hline
$4.0\times10^4$ & 30, 60, 120                                                     & 25, 50, Filling    & 3.0, 6.0, 9.0  & 0.001, 0.005, 0.02                                                       & 0.05, 0.1, 0.3, 0.95 & \begin{tabular}[c]{@{}c@{}}Fallback \& \\ No Fallback\end{tabular}     & 486                                                                  \\ \hline
$1.0\times10^5$ & 30, 60, 120                                                     & 25, 50, Filling    & 3.0, 6.0, 9.0  & 0.001, 0.005, 0.02                                                       & 0.05, 0.1, 0.3, 0.95 & \begin{tabular}[c]{@{}c@{}}Fallback \&\\  No Fallback\end{tabular}     & 513                                                                  \\ \hline
$4.0\times10^5$ & 30, 60, 120                                                     & 25, 50, Filling    & 3.0, 6.0, 9.0  & 0.001                                                                    & 0.1, 0.3, 0.95       & \begin{tabular}[c]{@{}c@{}}Fallback \&\\  No Fallback\end{tabular}     & 166                                                                  \\ \hline
$7.0\times10^5$ & 30, 60, 120                                                     & 25, 50, Filling    & 3.0, 6.0, 9.0  & \begin{tabular}[c]{@{}c@{}}0.0002,0.001,0.005,\\ 0.006 0.02\end{tabular} & 0.1, 0.3, 0.95       & \begin{tabular}[c]{@{}c@{}}Fallback \& \\ No Fallback\end{tabular}     & 619                                                                  \\ \hline
$1.2\times10^6$ & 30, 60, 120                                                     & 25, 50, Filling    & 3.0, 6.0, 9.0  & 0.001                                                                    & 0.05, 0.1, 0.95      & \begin{tabular}[c]{@{}c@{}}Fallback \& \\ No Fallback\end{tabular}     & 164                                                                  \\ \hline
\end{tabular}
\end{center}
\vskip 0.2cm
 \textit{Notes:}
BH and NS kicks are the same \citep{Hobbsetal2005}, except the case of mass fallback \citep{Belczynskietal2002}. A two segmented initial mass function (IMF) as given by \citet{Kroupa2001} was used for all models, $f_b$ - binary fraction. If the binary fraction is equal to 0.95 then binary parameters are chosen according to \citet{Kroupa1995}, otherwise eccentricity distribution is thermal, mass ratio distribution is uniform and semi-major distribution is uniform in logarithm, between $2(R_1+R_2)$ and 100 AU. $N$ - initial number of stars and binaries, $R_t$ - tidal radius, $R_h$ - half-mass radius, $W_0$ - King model parameter, $Z$ - cluster metallicity, filling - tidally filling model. For each initial number of objects different combinations of parameters are used to generate the initial model.
\end{minipage}
\end{table*}}
\begin{figure*}
\vspace{-0.99cm}
\includegraphics[height=14.3cm,angle=0,width=14.3 cm]{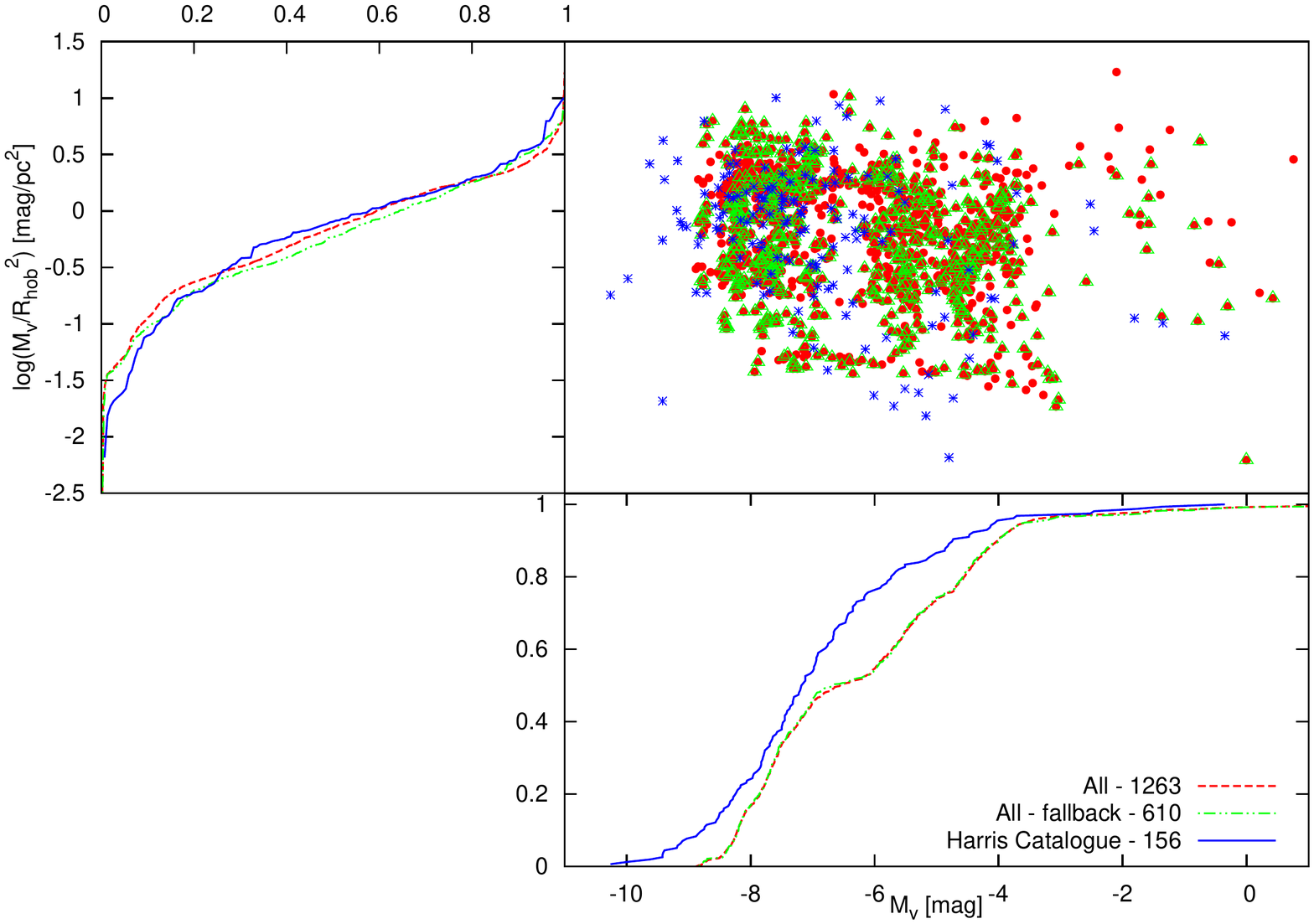}
\vspace{-1.3cm}
    \caption{The observational properties of Milky Way GCs (blue points) and modeled ones (at 12 Gyr) (red points). Green points represent simulated models in which mass fallback was enabled. The x-axis is the cluster absolute magnitude and y-axis is the average surface brightness inside the cluster half-light radius. Side figures show the distributions of the cluster absolute magnitude (bottom) and  the average surface brightness (left). Only cluster models that evolve up to at least 12 Gyrs are shown in the figure.} 
\label{fig:obs_prop}
  \end{figure*}
In this paper, we analysed thousands of star cluster models with different initial conditions that were simulated using the \textsc{mocca} code for star cluster simulation \citep{Gierszetal2013} as part of the \textsc{mocca} Survey I project. We extracted information for all coalescing BBHs that escaped or merged inside our cluster models. Using this data and the global properties of the simulated models, we compute local merger rate densities for BBHs originating from GCs. Having a large sample of BBHs with stellar mass and massive stellar BH components that will merge within a Hubble time from different type of cluster models, allows us to more accurately estimate merger rate densities. More detailed analysis of merging compact objects that could lead to GW emission from sources originating in GCs will be presented in subsequent papers.

\vspace{-0.45cm}  

\section{Method}\label{sec:met}

The \textsc{mocca} (MOnte Carlo Cluster simulAtor) code used for the star cluster simulations presented here is a numerical simulation code based on H\'{e}non's implementation of the Monte Carlo method \citep{Henon1971} to follow the evolution of stellar clusters. This method was substantially developed and improved by Stod\'{o}{\l}kiewicz in the early eighties \citep{Stodolkiewicz1986} and later by Giersz and his collaborators \citep[see][and reference therein for details about the \textsc{mocca} code and the Monte Carlo method]{Gierszetal2013}. \textsc{mocca} is a heterogeneous code, composed of independent modules that together model the entire cluster evolution on the level similar to state-of-the-art N-body models. It is able to follow most of the important physical processes that occur during the dynamical evolution of star clusters. In addition to the inclusion of relaxation, which drives the dynamical evolution of the system, \textsc{mocca} includes: (1) synthetic binary stellar evolution using the prescriptions provided by \citet{HurleyPT2000} and \citet{HurleyTP2002} (\textsc{bse} code), (2) direct integration procedures for small $N$ sub-systems using the \textsc{fewbody} code \citep{FCZR2004}, and (3) a realistic treatment of escape processes in tidally limited clusters based on \citet{FukushigeHeggie2000}.

\textsc{mocca} has been extensively tested against the results of N-body simulations of star cluster models comprising of thirty thousand to one million stars \citep[][and references therein]{Gierszetal2013,Heggie2014,Longetal2016,Mapelli2016}. The agreement between these two different types of simulations for both global cluster parameters and number of specific objects along with their properties is excellent. The output from \textsc{mocca} code is as detailed as direct N-body codes, but \textsc{mocca} is much faster. It needs about a day to complete the evolution of a realistic GC. The speed of the code makes it ideal for simulating a large set of models with varying initial conditions.




\vspace{-0.1cm}
\subsection{MOCCA-SURVEY Database I Models}\label{sec:mod}

In this section we describe the star cluster models that were simulated as part of the \textsc{mocca} Survey project. The simulated models are characterized by diverse parameters describing not only the initial global cluster properties, but also stellar and binary parameters. The parameters of these models are listed in Table \ref{tab:models}.

All models have a stellar initial mass function (IMF) given by \citet{Kroupa2001} with minimum and maximum stellar masses taken to be $0.08 M_{\odot}$ and $100.0 M_{\odot}$, respectively.
Supernovae (SN) natal kick velocities for neutron stars and BHs were drawn from a Maxwellian distribution with a dispersion of $265$ km/s \citep{Hobbsetal2005}. For most models, natal kicks for BHs were modified according to the mass fallback procedure described by \citet{Belczynskietal2002}. To model the Galactic potential, a point mass approximation with the Galaxy mass equal to the mass enclosed inside the cluster Galactocentric distance is assumed. Additionally, it is also assumed that all clusters have the same rotation velocity, equal to $220$ km/s. So, depending on the cluster mass and tidal radius the Galactocentric distances span from about 1 kpc to about 50 kpc. The number of models with different metallicities are as follows: 63, 831, 487, 64 and 503 for Z = 0.0002, 0.001, 0.005, 0.006 and 0.02, respectively. 

We would like to strongly stress that models for the Survey Database were not selected to match the observed Milky Way GCs. The sampling of the available parameter space was coarse, particularly for the total number of stars and half mass radius. Nevertheless, as it can be seen in Fig. \ref{fig:obs_prop}, except for few bright (massive and intermediate mass) Galactic GCs, the agreement with the observational properties of observed Galactic GCs \citep[][updated 2010]{Harris1996} and modeled ones that evolve up to at least 12 Gyrs is very good for the average surface brightness and rather poor for cluster absolute magnitude. We have to stress that any combination of global observational properties of GCs cannot be used to clearly distinguish between different cluster models because there is a strong degeneracy with respect to the initial conditions which is clearly seen in Fig. \ref{fig:obs_prop} in the case of the cluster absolute magnitude and the average surface brightness. Based on the aforementioned arguments, we can assume that the Survey cluster models are more or less representative of the GC population.

\vspace{-0.3cm}

\section{Merging BBHs from Cluster Models}\label{sec:numbers}

\begin{figure}
\vspace{-0.5cm}
\centering
  \includegraphics[height=7.0cm,width=7.0cm]{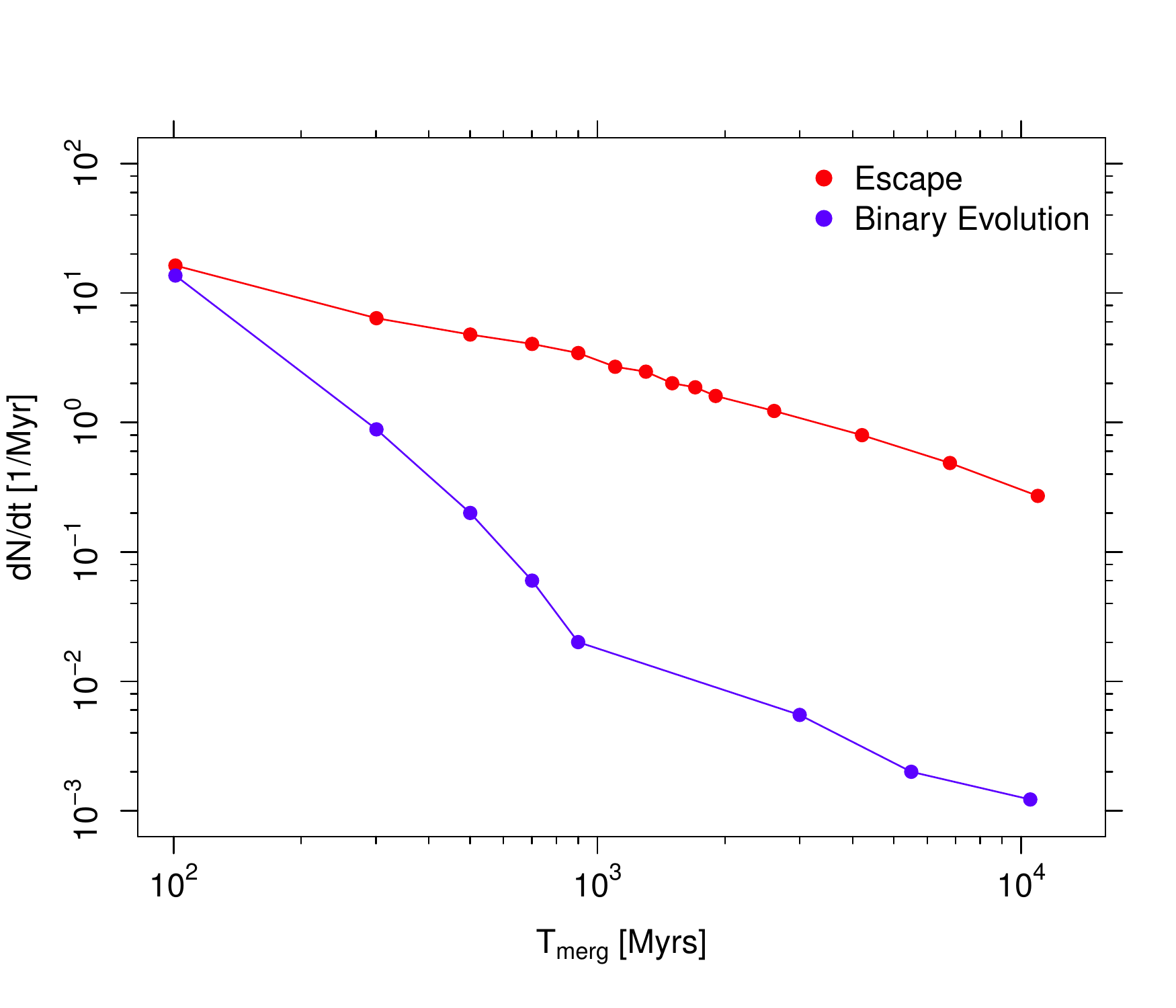}
  \vspace*{-0.5 cm}
  \captionof{figure}{Number of merging BBHs per unit time (1 Myr) as a function of merger time.
  Red points corresponds to BBHs that escaped from GC during the time of its evolution.  Blue points correspond to BBHs that merged inside the cluster due to GW emission.}
  \label{Fig:N_vs_Tmerg}
\end{figure}

\begin{figure}
\vspace{-0.2 cm}
  \includegraphics[height=8.0cm,width=7.0cm,angle=270]{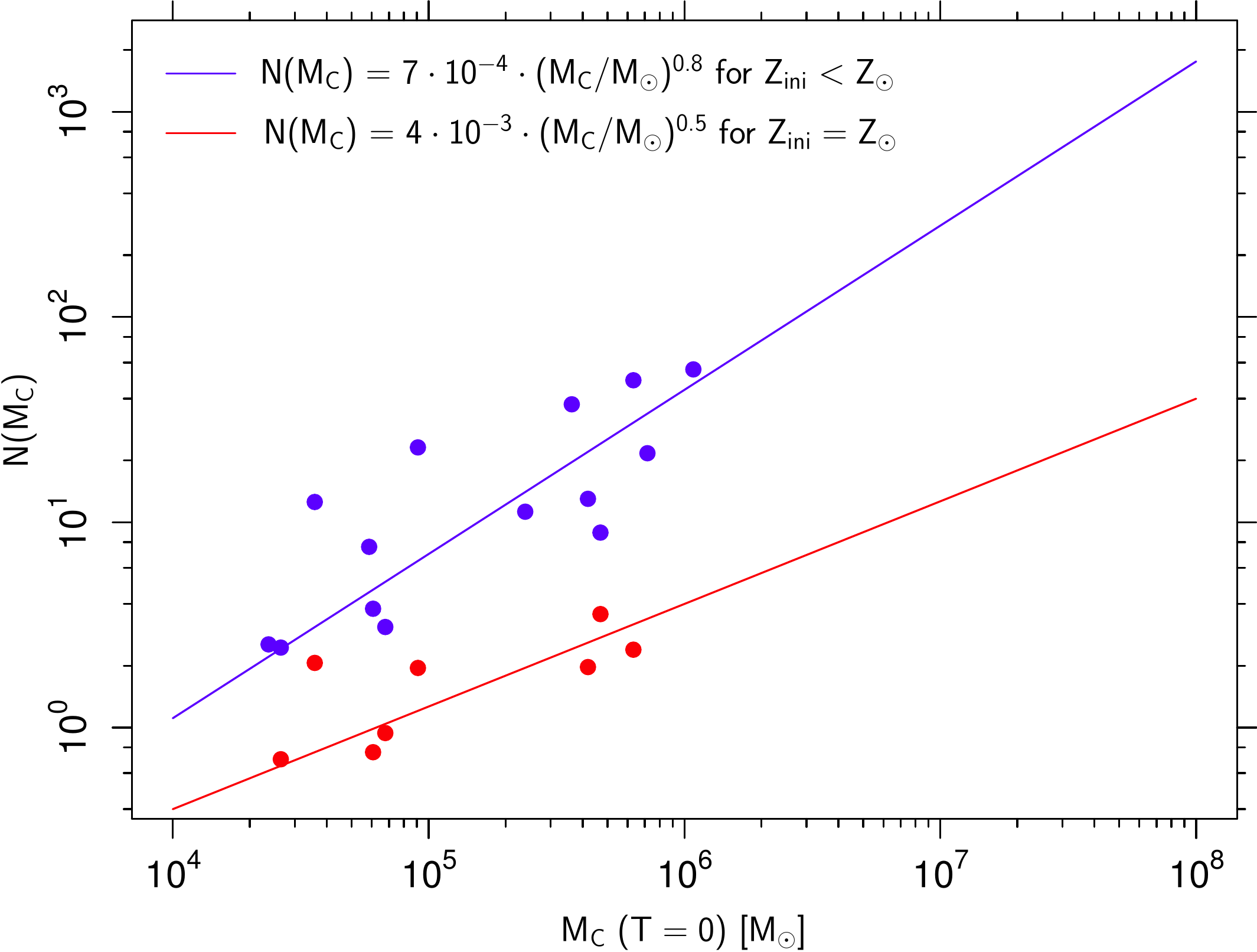}
  \vspace{-0.1 cm}
  \captionof{figure}{Normalized number of BBHs as a function of initial cluster mass $M_c$ with fitted function $N(M_c)$.  Data includes both escaped BBHs and BBHs that merge inside the cluster. Red and blue points correspond to two metallicities: solar ($Z = 0.02$) and sub-solar ($Z < 0.02$) respectively.}
  \label{Fig:N_vs_MGC}
  \vspace{-0.2 cm}
\end{figure}

Due to the extensive output provided by the \textsc{mocca} code, we have detailed information for all objects that escape from the system during its dynamical evolution. \textsc{mocca} also provides information for all binary evolution calls during the evolution of the cluster. We restricted our analysis to star cluster models in which BH natal kicks were computed according to the mass fallback prescription given by \citet{Belczynskietal2002}. In order to correctly determine the number of coalescing BBHs from 985 simulation models of Survey Database I in which mass fallback was enabled, we specifically searched each simulated model for all BBHs that escape the cluster model and go on to merge within a Hubble time and also for BBHs that merge inside the cluster via GW emission. For all escaping BBHs, we calculated the proper coalescence times using the formulae derived by \citet{Peters} and the time when the BBH escaped the GC. The number of merging BBHs are provided in Figures \ref{Fig:N_vs_Tmerg} and \ref{Fig:N_vs_MGC}.

Figure \ref{Fig:N_vs_Tmerg} shows the number of merging BBHs per unit time (1 Myr), as a function of merger time. Red and blue dots correspond to those BBHs that escape and merged outside the cluster model and those that merged inside the cluster, respectively.
Masses of both BHs in the BBH before merger were limited to 100 $M_{\odot}$. This was done to exclude merger events involving intermediate-mass black holes (IMBH) \citep{Gierszetal2015}.
From the simulation models, we found 15134 coalescing BBHs that escape the cluster and 3000 BBHs that merged inside the cluster.
Every merger event in both groups is limited to Hubble time - 13 Gyrs. 
The highest rate of mergers is during the early evolution of the GC model, for both escapers and GW mergers inside the cluster the rate is approximately the same. 
The difference gets higher during the evolution of the cluster. Due to many interactions, BBHs inside the cluster merge faster than those which escaped from it. Escaped BBHs do not undergo any dynamical perturbations outside the cluster and merge purely due to emission of GWs. 
During later times, the rate of merging BBHs decreases with time and is dominated by escaped BBHs. The slope for the number of escaped merging BBHs as a function merger time is -1 and for BBHs merging inside the cluster it is -2 (Figure \ref{Fig:N_vs_Tmerg}).  

Figure \ref{Fig:N_vs_MGC} shows the normalized number of mergers per simulation as a function of initial cluster mass. The normalization function is as follows:
\begin{equation}
N(M_c) = \frac {n} {n_s \cdot M_c/10^6 M_{\odot} }
\end{equation}
where $n$ is number of merging BBHs in $n_s$ simulations (number of simulations with a unique mass) and $M_c$ is GC mass.
Red points corresponds to BBH mergers from simulations with solar metallicity $Z = 0.02$, and blue points corresponds to BBH mergers from simulations with smaller metallicities $Z < 0.02$
The total number of mergers events are 18134, in that for $Z = 0.02$ - $865$ mergers, and for $Z < 0.02$ - $17269$ mergers.
The theoretical models were fitted to the data:
\begin{equation}
Z = 0.02 \rightarrow N(M_c) = 4 \cdot 10^{-3} \cdot ({M_c/M_{\odot}})^{0.5}
\end{equation}
\vspace{-0.36 cm}
\begin{equation}
Z < 0.02 \rightarrow N(M_c) = 7 \cdot 10^{-4} \cdot ({M_c/M_{\odot}})^{0.8}
\end{equation}
 
Regardless of the metallicity, if the mass of a GC model is large then the number of merging BBHs is higher. The difference occurs only in the value of the rate, low metallicity clusters have a greater ratio of producing merging BBHs compared to higher metallicity cluster models. \citet{Rodriguez1} obtained a linear expression for the number of merging BBHs against the cluster mass and got 50\% larger number of mergers compared to our data. This maybe because the maximum initial stellar mass in their models was $150 M_{\odot}$ compared to $100 M_{\odot}$ in our models. Including more massive stars initially will produce more coalescing BBHs in GC models.

\vspace{-0.8cm}
\section{Local rate density  of BBH mergers}\label{sec:rates}

The local merger rate density calculation will follow the 
same formalism as used in calculation of the local merger rate in the 
case of field evolution \cite{2004A&A...415..407B}.

Let us denote the GC star formation rate as
$SFR_{GC}(z)$, and the average stellar mass in a cluster as 
$M_{av}$. The probability of formation of a BBH
in a cluster per unit solar mass is denoted as $P_{BBH}$.
The BBH have different properties such as 
the delay time from the formation until the merger $t_{del}$, 
and the chirp mass $ {\cal M} = (m_1 m_2)^{3/5}(m_1 +m_2)^{-1/5}$.
It is more convenient to express the probability of formation of BBHs as a differential probability per unit delay time and per unit
chirp mass: $\frac{d P_{BBH}}{d t_{del}  d {\cal M}}$.
Given these quantities the local merger rate density can be expressed
as:
\begin{equation}
\frac{d{\cal R}}{d {\cal M}} = \int_0^{T_{Hubble}} d t_{del} \frac{SFR_{GC}(z(t_{del}))}{M_{av}} 
\frac{d P_{BBH}}{d t_{del}  d {\cal M}}
\label{eq:rate}
\end{equation}
where $z(t_{del})$ is the redshift corresponding to the cosmic time $t$.
We use the standard $\Lambda$ CDM cosmology with the Hubble constant 
$H_0=72$km\,s$^{-1}$\,Mpc$^{-1}$.  The total local merger rate of BBHs
originating in clusters is then 
\begin{equation}
{\cal R} = \int {d{\cal R} \over d {\cal M}}  d {\cal M}.
\end{equation}

Given a numerical simulation of clusters we can discretize equation \ref{eq:rate}.
Let us consider a simulation of clusters with a total mass $M_{sim}$ in which 
$N$ BBHs were formed with the delay times $t_{del}^i$ and chirp 
masses ${\cal M}^i$. We can approximate the probability of formation of BBHs as
\begin{equation}
{d P_{BBH}\over  d t_{del}  d {\cal M}} = M_{sim}^{-1} \sum_{i=1}^N \delta(t_{del}-t_{del}^i) (\delta{{\cal M} - \cal M}^i)
\label{eq:numprob}
\end{equation}
Inserting equation \ref{eq:numprob} into \ref{eq:rate} we obtain
\begin{equation}
{d{\cal R} \over d {\cal M}} = M_{sim}^{-1} \sum_{i=1}^N \frac{SFR_{GC}(z(t_{del}^i))}{M_{av}} .
\end{equation}

The GC star formation rate has been estimated by \citet{2013MNRAS.432.3250K}.
We are using their result obtained with the inclusion of constraints from low redshifts, see the 
thick line in the top panel of Figure 6 in \citet{2013MNRAS.432.3250K}.
We are  using the results of the \textsc{mocca} Survey I simulations 
with fallback, which is a total of 985 clusters with the total mass simulated 
$M_{sim}= 2.82\times 10^8{\rm M}_\odot$, we obtain the local merger rate density of GC originating BBHs:
\begin{equation}
{\cal R} = 5.4\:{\rm Gpc}^{-3}\,{\rm yr}^{-1}
\end{equation}

\begin{figure}
\vspace{-1.61cm}
\includegraphics[height=9.6cm,width=7.9cm]{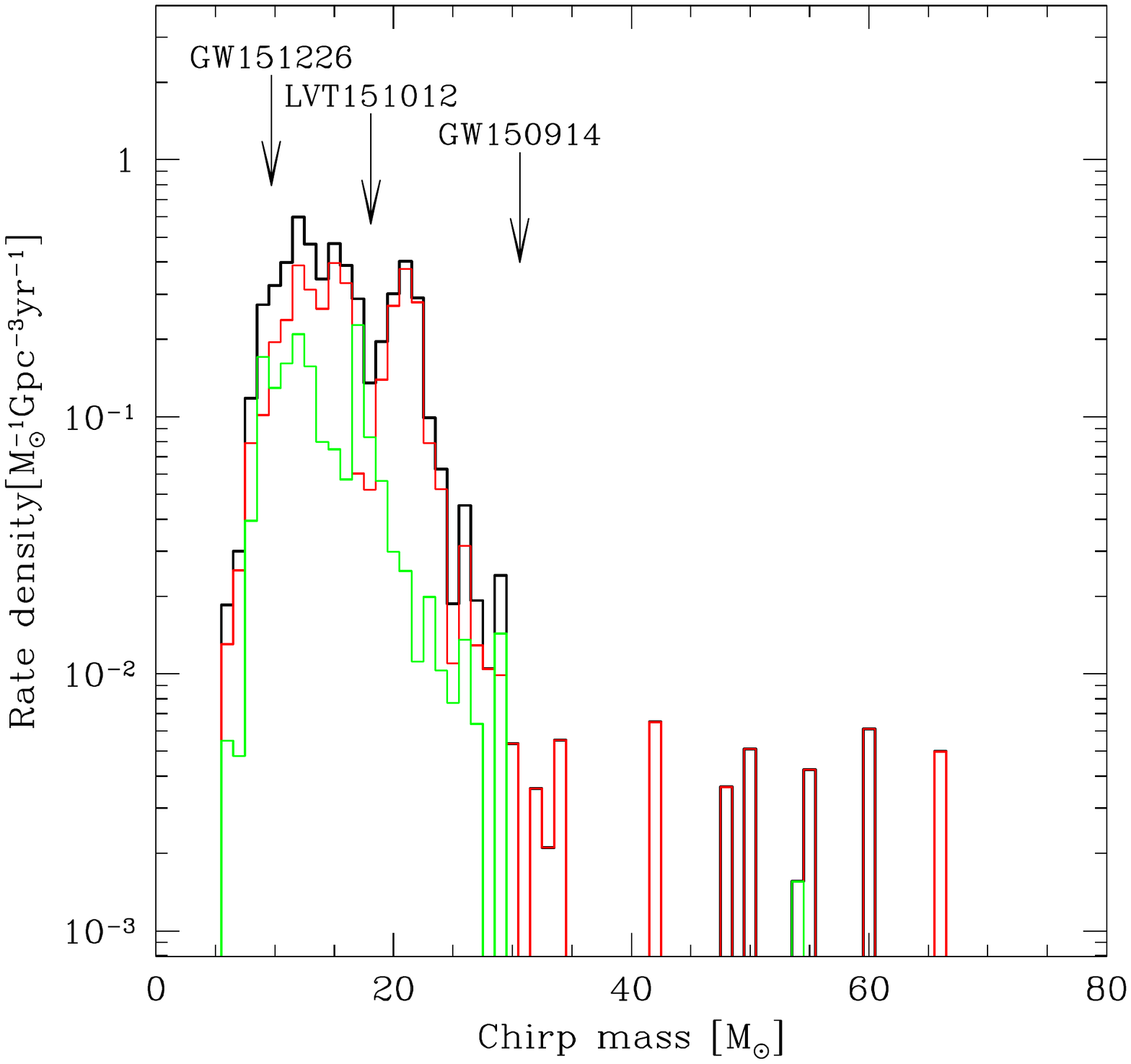}
\vspace*{-2.3cm}
\caption{The figure shows the differential rate density per unit chirp mass of coalescing BBHs. The dark line shows the total merger rate density for all models. The red line corresponds to total merger rate density for models with sub-solar metallicities and the green line shows it for solar metallicity models. The chirp masses of GW events detected by LIGO are marked by arrows.}
\label{fig:densi}
\vspace{-0.3cm}
\end{figure}

In Figure \ref{fig:densi} we present the differential rate density per unit chirp mass. The rate is dominated by sub-solar metallicity models and is highest for chirp masses 10-30 $M_{\odot}$ with a tail up to 70 $M_{\odot}$. 
This result is consistent with LIGO observations.The chirp mass distribution for these BBHs is related to the masses of BHs in the simulated models. BH masses strongly depend on the initial mass and the stellar evolution (approximate prescriptions provided by \textsc{bse} code) of BH progenitors.
We must note that the rate density we have calculated is very conservative.
The models of GCs we have used did not include the highest mass GCs,
as the maximum mass of a simulated cluster was $10^6$ $M_\odot$.
In order to assess the influence of more massive clusters we have introduced a
BBH production rate efficiency, $N(M_{c})$, the number of merging BBH systems per million solar masses of stars. We plot this quantity as a function of
the cluster mass in Figure~\ref{Fig:N_vs_MGC}. While the scatter in the plot is quite 
significant, one can see a rough trend of the increase in the number of BBH per unit 
star forming mass with the cluster mass. The scaling is approximately as 
$N(M_{c}) \approx M^{0.5 - 0.8}$.  Thus inclusion of the more massive GCs with 
mass of $10^7$M$_\odot$ 
will likely increase the local rate by a factor of 3-5 (which comes from the scaling $10^{0.5}$ to $10^{0.8}$).
Additionally, the uncertainty in the metallicity composition of GCs in early galaxies and the uncertainties connected with stellar IMF and the maximum stellar mass may also introduce an additional increase in the merger rate. Note that the \textsc{mocca} Survey I 
simulations are reproducing the current population of GCs in  the Milky Way, 
while the initial population of GCs might have contained more clusters with low metallicity 
that have dissolved by today. 

A simple calculation can be used to  the expected rate of events in the first LIGO observing run (O1). 
In the low mass regime when ${\cal M} < 100$M$_\odot$ one can assume that in O1 the sensitivity distance scales as
\vspace{-0.3cm}
\begin{equation}
 D_{mean}({\cal M}) = D_{mean}(1.2 {\rm M}_\odot) \left(\frac{{\cal M}}{1.2 M_\odot}\right)^{5/6}
 \end{equation}
 where $D_{mean}(1.2 {\rm M}_\odot) \approx 80$Mpc. 
 If we additionally assume that the rate density weakly depends on the distance
(or redshift) we obtain the expected number of detections
\begin{equation}
N_{O1} = T_{O1} \int {d{\cal R} \over d {\cal M}} \frac{4\pi D_{mean}^3({\cal M)} }{3} d {\cal M}
\end{equation}
inserting the result of our simulations we obtain the expected number of O1 
detections $N_{O1} \approx 0.36$. However, allowing for the contribution of the most massive 
GC this number can be increased by a factor of $3-5$ to reach approximately $1.7$ expected detections in O1.

\section{Discussion \& Conclusion}\label{sec:conc} 

Using the results from \textsc{mocca} Survey I models with mass fallback enabled, we extracted data for BBHs that escape from or merge inside the GC models. We use this data to calculate local merger rate density of $5.4$ ${\rm Gpc}^{-3}\,{\rm yr}^{-1}$. The merger rate density computed through our analysis is comparable to the local merger rate density of $\sim 5$ ${\rm Gpc}^{-3}\,{\rm yr}^{-1}$ calculated by \citet{Rodriguez2}. The simulation models used by \citet{Rodriguez2} had higher initial maximum star mass compared to our simulation models. Including GC models representative of most massive observed GCs can increase our calculated merger rate by a factor of $3-5$. Another difference between our results and those of \citet{Rodriguez2} is that only about 2/3 of our 985 cluster models that were used to calculate the average GC mass for the rate density calculation contributed BBHs that would merge within a Hubble time and in which coalescing BHs had mass less than 100 $M_{\odot}$. Merger rate density would increase if we only used the average mass of the models which contributed observable stellar mass coalescing BBHs. Additionally, uncertainties associated with the stellar IMF and evolution of most massive stars may also increase the rate and influence the chirp mass distribution of merging BBHs. 

The local rate density we have obtained
is  compatible with the lower bound on the
observed LIGO BBH merger rate density of  9-240 ${\rm Gpc}^{-3}\,{\rm yr}^{-1}$ \citep{Ligo2016}. The GC originating BBH rate density is a product of two factors. On the one hand the
amount of stellar mass in GCs is much smaller than that in the
field, and on the other the efficiency of production
of BBH in GCs is higher. Additionally, stars were formed in GCs in the early Universe, with the maximum at $z>2$,
and currently there is very little star formation in GCs.
On the other hand the field star formation is still taking place in galaxies.
Thus the main advantage of the GC formation
is the increased BBH formation efficiency due to many body interactions. The rate density we have obtained is on the lower end of the
value allowed by observations.  The rate density of the field binaries
can be much higher - see e.g. \citet{KB2016}.
The field evolution models have more uncertainties and can encompass a broader range of values.
With the second observing run of LIGO, we expect that the accuracy of the rate density estimate will be better. 
and this can also be verified by future LIGO runs.
The rate density we find can be increased due to a few factors: the inclusion
of more massive clusters, where the BBH formation efficiency is higher; changing the metallicity composition of the GCs star formation; and increasing maximum stellar mass.
These considerations can allow for an increase of the calculated rate by a factor of up to 5. This would result in few tens of BBH merger events per year and this is consistent with the results from \citet{ant2}. We conclude that the merger rate density of BBHs with stellar mass and massive stellar BH components originating from GCs would be not more than about 30 ${\rm Gpc}^{-3}\,{\rm yr}^{-1}$. It is quite likely that the observed population of coalescing BBHs consists of
objects with different formation histories.

\vspace{-0.6cm}
\section*{Acknowledgements}\label{sec:ack}  

We would like to thank the referee and Daniel Holz for their comments and suggestions. AA and MG were supported by the Polish National Science Centre (PNSC) through the grant DEC-2012/07/B/ST9/04412. AA would also like to acknowledge support by the PNSC through the grant UMO-2015/17/N/ST9/02573. DGR and MS were supported by the POMOST/2012-6/11 Program of FNP co-financed by the European Regional Development Fund and by PNSC grant UMO-2014/14/M/ST9/00707. TB acknowledges support from PNSC through the grant UMO-2014/15/Z/ST9/00038.

\bsp
\label{lastpage}

\end{document}